\documentstyle[12pt,epsfig]{article}
\def\bge{\begin{equation}}
\def\ene{\end{equation}}
\def\bg{\begin{eqnarray}}
\def\en{\end{eqnarray}}
\def\nn{\nonumber}

\def\S0{{\Sigma^0}}

\def\k0bar{\bar{K}^0}

\begin{document}
\renewcommand{\thefootnote}{\fnsymbol{footnote}}
\begin{flushright}
ADP-98-8/T287
\end{flushright}
\begin{center}
{\LARGE $\sigma$ and $\omega$ meson propagation in a dense nuclear medium}
\end{center}
\vspace{0.5cm}
\begin{center}
\begin{large}
K.~Saito\footnote{ksaito@nucl.phys.tohoku.ac.jp} \\
Physics Division, Tohoku College of Pharmacy \\
Sendai 981-8558, Japan \\
K.~Tsushima\footnote{ktsushim@physics.adelaide.edu.au}, 
A.W.~Thomas\footnote{athomas@physics.adelaide.edu.au} 
and A.G.~Williams\footnote{awilliam@physics.adelaide.edu.au} \\
Special Research Center for the Subatomic Structure of Matter \\
and Department of Physics and Mathematical Physics \\
The University of Adelaide, SA 5005, Australia 
\end{large}
\end{center}
%
\newpage
\begin{abstract}
The propagation of the scalar ($\sigma$) and vector ($\omega$) mesons 
in nuclear matter is studied in detail using the Walecka model  
over a wide range of densities and including the effects of a finite 
$\sigma$ width through the inclusion of a two-pion loop.  We calculate 
the dispersion relation and spectral functions of the $\sigma$ and 
(transverse and longitudinal) $\omega$ mesons, including the effect of 
$\sigma$-$\omega$ mixing in matter.  It is shown that the mixing effect 
is quite important in the propagation of the (longitudinal) $\omega$ 
and $\sigma$ mesons above normal nuclear matter density.  
We find that there is a 
two-peak structure in the spectral function of the $\sigma$ channel, 
caused by $\sigma$-$\omega$ mixing.  
\end{abstract}
\vspace{0.5cm}
\newpage

Recently much attention has been focussed on the variation 
of hadron properties in hot and/or dense nuclear matter\cite{qm97}.  
In particular, 
the medium modification of the light vector ($\rho$, $\omega$ and 
$\phi$) meson masses has been investigated by many 
authors\cite{qhd,will,br,asa,qcd,saito,fri,rec}.  
Experimentally, dilepton production in hot and/or dense nuclear matter 
produced by relativistic heavy ion collisions can provide a unique 
tool to measure such modifications.  It is well known that data 
obtained at the CERN/SPS by the CERES\cite{ceres} and HELIOS\cite{hel} 
collaborations  
show a significant amount of strength below the free $\rho$ 
meson peak.  Some authors\cite{li} have concluded that this 
is caused by a 
downward shift of the rho meson mass in dense nuclear matter.  
To test this idea, measurements of the dilepton spectrum from vector 
mesons produced in nuclei are planned at TJNAF\cite{jlab} and 
GSI\cite{gsi} (see also Ref.\cite{ins}).  

On the other hand, the $\sigma$ meson has been treated 
as a correlated two-pion state in the scalar channel for a 
long time.  However, recently some people have reanalysed the 
s-wave (I=0) $\pi\pi$-scattering phase shift, and argued  
for the existence of the $\sigma$ as a genuine resonance\cite{sigma}.  
{}From the theoretical side, 
it is also anticipated that the $\sigma$ meson may play an important 
role as the chiral partner of the $\pi$ meson\cite{kuni}. 
Some experimental possibilities have been proposed  
for investigating the behaviour of the $\sigma$ in hot and/or 
dense nuclear matter\cite{kuni,shim}.

It is well known that in nuclear matter the $\sigma$ can 
couple to the longitudinal mode of the $\omega$ meson\cite{chin}.  
This work is an extension and elaboration of studies in Ref.\cite{will} 
where the invariant mass of the $\omega$ meson moving 
in nuclear matter was studied.  
Here, we generalize this work by including free space widths for the $\sigma$
and $\omega$ and by studying $\sigma$-$\omega$ mixing in some detail as 
a function of density. Recently Wolf 
{\it et al}.\cite{wolf} studied the effect of $\sigma$-$\omega$ 
mixing on the $e^+e^-$-pair production in relativistic heavy ion 
collisions.  However they calculated only the lowest order  
mixing diagram, and did not include the nucleon-loop diagrams for 
the $\sigma$ and $\omega$ meson propagators.  
The interesting results obtained in these earlier papers and the 
experimental relevance of the problem suggests that it is time for 
a more complete self-consistent calculation.

A complete investigation of the propagation of $\sigma$ and $\omega$ 
mesons in a dense nuclear medium requires that one includes 
the effect of $\sigma$-$\omega$ mixing.  In this 
letter we will study in detail the medium modification of $\sigma$ and 
$\omega$ mesons in dense (symmetric) nuclear matter using the 
simplest version of quantum hadrodynamics (QHD or the 
Walecka model)\cite{sw} and including $\sigma$-$\omega$ mixing, 
$\sigma \pi \pi$ coupling, and the $\omega$ width.  

The lagrangian density of the Walecka model (QHD-I) is well known 
and a full description of it can be found in Ref.\cite{sw}.  
We first treat the 
nuclear ground state in relativistic Hartree approximation (RHA),  
which is also explained in Ref.\cite{sw}.    
Here we shall ignore the coupling to the channels involving 
iso-vector mesons, neglecting, 
for example, $\omega + N \to \pi + N$ and 
$\omega + N \to \rho + N$\cite{iv}.  

To compute 
the meson propagators, we sum over the ring diagrams, which consist 
of repeated insertions of the lowest order one-loop proper 
polarization part.  This is the relativistic, random phase 
approximation (RPA)\cite{chin}.  Since we want 
to include $\sigma$-$\omega$ mixing, it is convenient to use 
a meson propagator ${\cal D}_{ab}$ in the form of a $5 \times 5$ matrix 
with indices $a$, $b$ running from 0 to 4, where 4 is for the 
$\sigma$ channel and 0 $\sim$ 3 are for the $\omega$.  

Dyson's equation for the full propagator, ${\cal D}$, is given in 
matrix form as: 
\bge
{\cal D} = {\cal D}^0 + {\cal D}^0 \Pi {\cal D},  \label{full}
\ene
where ${\cal D}^0$ is the lowest order meson propagator given by 
a block-diagonal form as 
\bge
{\cal D}^0 = \left( 
\begin{array}{cc}
    D^0_{\mu \nu} & 0 \\
    0 & \Delta_0      
\end{array} 
\right). 
\label{free}
\ene
In Eq.(\ref{free}) the noninteracting propagators for the $\sigma$ 
and $\omega$ are given respectively by 
\bg
\Delta_0(q) &=& \frac{1}{q_\mu^2 - m_\sigma^2 + i\epsilon},  
\label{frees} \\
D^0_{\mu \nu}(q) &=& \frac{\xi_{\mu \nu}}{
  q_\mu^2 - m_\omega^2 + i\epsilon}, \label{freeo}
\en
where $\xi_{\mu \nu} = - g_{\mu \nu} + (q_\mu q_\nu /q_\mu^2)$, 
$q_\mu^2 = q_0^2 - |{\vec q}|^2$, and $m_\sigma$ and $m_\omega$ 
are the free $\sigma$ and $\omega$ meson masses.  

The polarization insertion in Eq.(\ref{full}) is also then given by 
a $5 \times 5$ matrix, 
\bge
\Pi = \left( 
\begin{array}{cc}
    \Pi_{\mu \nu}(q) & \Pi_\nu^m(q) \\
    \Pi_\mu^m(q) & \Pi_s(q)      
\end{array} 
\right),
\label{pol}
\ene
where the lowest order scalar, vector and scalar-vector-mixed 
polarization insertions are given respectively as 
\bg
\Pi_s(q) &=& -ig_s^2 \int \frac{d^4k}{(2\pi)^4}\mbox{Tr}[G(k)G(k+q)] 
     + \frac{3}{2} ig_{\sigma\pi}^2 m_\pi^2 \int 
     \frac{d^4k}{(2\pi)^4} \Delta_\pi(k) \Delta_\pi(k+q), 
\label{pis} \\
\Pi_{\mu \nu}(q) &=& -ig_v^2 \int \frac{d^4k}{(2\pi)^4}
     \mbox{Tr}[G(k) \gamma_\mu G(k+q) \gamma_\nu], \label{piv} \\  
\Pi_\mu^m(q) &=& ig_sg_v \int \frac{d^4k}{(2\pi)^4}
     \mbox{Tr}[G(k) \gamma_\mu G(k+q)]. \label{pim}
\en
Note that we have added the contribution from the pion-loop to $\Pi_s$ (the 
second term in the r.h.s. of Eq.(\ref{pis})) in order to treat 
the $\sigma$ more realistically.  
The pion propagator, $\Delta_\pi$, is given 
by equation (\ref{frees}) with the pion mass $m_\pi$ instead of 
$m_\sigma$.  Here $g_v$, $g_s$ and $g_{\sigma\pi}$ are respectively the 
nucleon-$\omega$, nucleon-$\sigma$ and 
$\sigma$-$\pi$ coupling constants.  We denote 
$G(k)$ as the self-consistent RHA nucleon propagator, which is  
given by the sum of the Feynman (F) part and 
the density-dependent (D) part as 
\bg
G(k) &=& G_F(k) + G_D(k), \nn \\
   &=& (\gamma^\mu k^*_\mu + M^*) \left[ 
   \frac{1}{k^{*2}_\mu - M^{*2} + i\epsilon} + \frac{i\pi}{E^*_k} 
   \delta(k^*_0 - E_k^*) \theta(k_F - |{\vec k}|) \right], 
\label{propn}
\en
where $k^{*\mu}=(k^0 - g_vV^0, {\vec k})$ ($V^0$ is the mean value of 
the $\omega$ field), $E_k^*=\sqrt{{\vec k}^2 + M^{*2}}$ ($M^*$ is 
the effective nucleon mass in matter) and $k_F$ is the Fermi momentum.  
Therefore, each polarization insertion (except the pion-loop 
contribution) can be divided into two pieces: the Feynman (F) 
(or vacuum) part, which does not involve 
the $\theta(k_F - |{\vec k}|)$ term, and the 
density-dependent (D) part.  The explicit expressions for the various 
components of D can be found in Ref.\cite{lim}. 

We can work out the F parts using the method of dimensional 
regularization to remove the divergence in the loop calculations. 
We show the results explicitly.  For the $\sigma$, we must renormalize 
two terms: the nucleon-loop and pion-loop diagrams.  For the 
nucleon-loop contribution to the $\sigma$ 
we introduce the usual counter terms to the lagrangian\cite{sw} 
\bge
\delta {\cal L}_\sigma = \sum_{l=2}^4 \frac{\alpha_l}{l!} 
\sigma^l + \frac{\zeta}{2} (\partial \sigma)^2, \label{conts}
\ene
which includes quadratic, cubic, quartic and wavefunction renormalization.
To get the ``physical'' properties of the $\sigma$ meson in free space, 
we impose the following condition on the F part of the nucleon-loop 
diagram\cite{suzuki}: 
\bge
\Pi_s^{N-loop}(q_\mu^2, M^*=M) = 
\frac{\partial}{\partial q_\mu^2} \Pi_s^{N-loop}(q_\mu^2, M^*=M) = 0 
\ \ \ \mbox{ at } q_\mu^2 = m_\sigma^2,  \label{scond}
\ene
where $M$ is the free nucleon mass.  We adjust the 
coefficients $\alpha_2$ and $\zeta$ to satisfy this condition.  For 
$\alpha_3$ and $\alpha_4$ we use the usual values given in 
Ref.\cite{sw}.  For the pion-loop, we also introduce an appropriate 
counter term to the lagrangian, 
and require a similar condition to Eq.(\ref{scond}) for 
the polarization insertion 
of the pion-loop diagram in free space: 
\bge
{\Re}e\Pi_s^{\pi-loop}(q_\mu^2) = 0 
\ \ \ \mbox{ at } q_\mu^2 = m_\sigma^2.  \label{picond}
\ene
Finally, we find 
\bg
\Pi_s^{N-loop}(q) &=& \frac{3g_s^2}{2\pi^2} \biggl[ 
\frac{1}{6} (m_\sigma^2 - q_\mu^2)  \\ \nn
&-& \left( M^{*2} - \frac{q_\mu^2}{6} \right) \left( 2 
\ln\frac{M^*}{M} 
+ f(x_q) -  f(z_s) \right)  \\  \nn
&+& \frac{q_\mu^2}{3} \left( \frac{M^{*2}}{q_\mu^2} 
(f(x_q)-2) - \frac{M^2}{m_\sigma^2} 
(f(z_s)-2) \right) \\ \nn
&-& (M^{*2} - M^2) ( f(z_s)-2) 
+ 2M(M^*-M) + 3(M^*-M)^2 \biggr],  \label{n-loop}
\en
where $x_q=1-\frac{4M^{*2}}{q_\mu^2}$, $z_s=1-\frac{4M^2}{m_\sigma^2}$ 
and 
\bge
\Pi_s^{\pi-loop}(q) = \frac{3g_{\sigma\pi}^2}{32\pi^2} m_\pi^2 
\left[ f(x_\pi) - {\Re}ef(z_\pi) \right], \label{pi-loop}
\ene
where $x_\pi=1-\frac{4m_\pi^2}{q_\mu^2}$ and 
$z_\pi=1-\frac{4m_\pi^2}{m_\sigma^2}$, and 
\bge
f(y) = \left\{ 
\begin{array}{rl}
\sqrt{y} \ln\frac{\sqrt{y}+1}{\sqrt{y}-1}, & \ \ \ 
\mbox{ for $1 \leq y < +\infty$} \\
\sqrt{y} \ln\frac{1+\sqrt{y}}{1-\sqrt{y}} -i \pi \sqrt{y}, & \ \ \
\mbox{ for $0 < y < 1$} \\ 
2\sqrt{-y} \tan^{-1}\frac{1}{\sqrt{-y}}. & \ \ \
\mbox{ for $y \leq 0$} \end{array} \right.  \label{fff}
\ene

For the $\omega$, we add the counter term for the wavefunction 
renormalization to the lagrangian and require the 
following condition for the 
F part of the polarization insertion of the nucleon-loop: 
\bge
\Pi_{\mu \nu}^{N-loop}(q) = \xi_{\mu \nu} \Pi^{N-loop}(q_\mu^2, 
M^*=M) = 0 \ \ \ \mbox{ at } q_\mu^2 = m_\omega^2.  \label{vcond}
\ene
Then, we get 
\bg
\Pi^{N-loop}(q) &=& \frac{g_v^2}{6\pi^2} q_\mu^2 \biggl[ 
2 \ln\frac{M^*}{M} - 4\left( \frac{M^{*2}}{q_\mu^2} - 
\frac{M^2}{m_\omega^2} \right)  \\ \nn
&+& \left( 1 + 2\frac{M^{*2}}{q_\mu^2} \right) 
f(x_q) - 
\left( 1 + 2\frac{M^2}{m_\omega^2} \right) 
f(z_v) \biggr],  \label{v-loop}
\en
where $z_v=1-\frac{4M^2}{m_\omega^2}$.
Note that vacuum fluctuations do not contribute 
to the mixed part of Eq.(\ref{pim}).  

Using these polarization insertions, we define the dielectric function 
$\epsilon$ as\cite{chin,sw,lim}
\bg
\epsilon &=& \mbox{det}(1 - {\cal D}^0 \Pi), \nn \\
   &=& \epsilon_T^2 \times \epsilon_{SL}, \label{diel}
\en
where $\epsilon_{T[SL]}$ is the dielectric function for the 
transverse (T) [scalar and longitudinal (SL)] mode.  
(Note that the full propagator Eq.(\ref{full}) can be rewritten 
as ${\cal D}={\cal D}^0/(1-{\cal D}^0\Pi)$ and hence 
$\mbox{det}{\cal D} = \mbox{det}{\cal D}^0/\epsilon$.)  
Then, we find 
\bg
\epsilon_T &=& 1 - d_0 \Pi_T, \label{trns} \\
\epsilon_{SL} &=& (1 - d_0 \Pi_L)(1 - \Delta_0 \Pi_s) - 
\frac{q_\mu^2}{q^2} \Delta_0 d_0 (\Pi_0^m)^2. \label{long}
\en
Here $q=|{\vec q}|$ and $d_0^{-1}= q_\mu^2 - m_\omega^2 + 
im_\omega\Gamma_\omega^0$, where we add the width of the $\omega$ 
in free space ($\Gamma_\omega^0=9.8$ MeV).  
The transverse ($\Pi_T$) and longitudinal 
($\Pi_L$) components of the polarization insertion Eq.(\ref{piv}) 
are respectively 
defined by $\Pi_{11}$ (or $\Pi_{22}$) and  $\Pi_{33}-\Pi_{00}$ 
(we choose the direction of ${\vec q}$ as the $z$-axis).  
Note that $\Pi_0^m$ is the 0-th component of the mixed polarization 
insertion Eq.(\ref{pim}), which vanishes when $q=0$ (because of 
current conservation).  
The eigencondition for determining the collective excitation spectrum is 
just equivalent to searching for the zeros of the dielectric functions. 
Since we are interested in the medium modification of the meson 
propagation, we restrict ourselves here to the meson branch in 
the time-like region.  

To study the dispersion relation of the meson branch, we first have 
to solve the nuclear ground state within RHA.  Because we choose 
$\alpha_2$ and $\zeta$ to satisfy 
the renormalization condition for the $\sigma$ at $q_\mu^2=m_\sigma^2$ 
(see Eq.(\ref{scond})), the 
total energy density is written as 
\bge
{\cal E} = {\cal E}_0 + \frac{1}{2\pi^2}M^2(M-M^*)^2 \left[ 
  \frac{m_\sigma^2}{4M^2} + \frac{3}{2} f(z_s) 
-3 \right], \label{energy} 
\ene
where ${\cal E}_0$ is the usual one (in RHA) given in Ref.\cite{sw}.  
(Note that in Ref.\cite{sw} the renormalization condition on nucleon
loops is imposed at $q_\mu^2$=0.
This difference yields the second term of 
the r.h.s. of Eq.(\ref{energy})\cite{suzuki}. As a result, our 
model gives the same physical quantities as those of Ref.\cite{sw}.) 
Requiring 
the saturation condition, ${\cal E}/\rho_B - M = - 15.7$ MeV at 
$\rho_0=0.15$ fm$^{-3}$ (where $\rho_B$ and $\rho_0$ are respectively 
the density of nuclear matter and the saturation density), we  
determine the coupling constants $g_s^2$ and $g_v^2$: $g_s^2=101.50$ 
and $g_v^2=72.296$.  In the calculation we fix the values of the 
free hadron masses to be $M=939$ MeV, $m_\sigma=550$ MeV, $m_\omega=783$ MeV 
and $m_\pi=138$ MeV. 
This yields the effective nucleon mass $M^*/M=0.73$ at $\rho_0$ 
and the incompressibility $K=452$ MeV.  We do not consider the 
possibility of medium modification of the pion loop or any 
of the coupling constants in the present work.  

To determine the coupling constant $g_{\sigma\pi}$, we adjust it to 
fix the width of the $\sigma$ in free space.  Since the full propagator 
of $\sigma$ in free space is given in terms of the pion-loop 
polarization insertion, the width $\Gamma_\sigma^0$ is given by 
\bge
\Gamma_\sigma^0 = - \frac{{\Im}m\Pi_s^{\pi-loop}}{m_\sigma} 
 \ \ \ \mbox{ at } q_\mu^2 = m_\sigma^2.  \label{swidth}
\ene
If we choose $\Gamma_\sigma^0=$ 300 MeV\cite{sigma}, 
we find $g_{\sigma\pi}=18.33$.  

\begin{figure}[htb]
\begin{center}
\epsfig{file=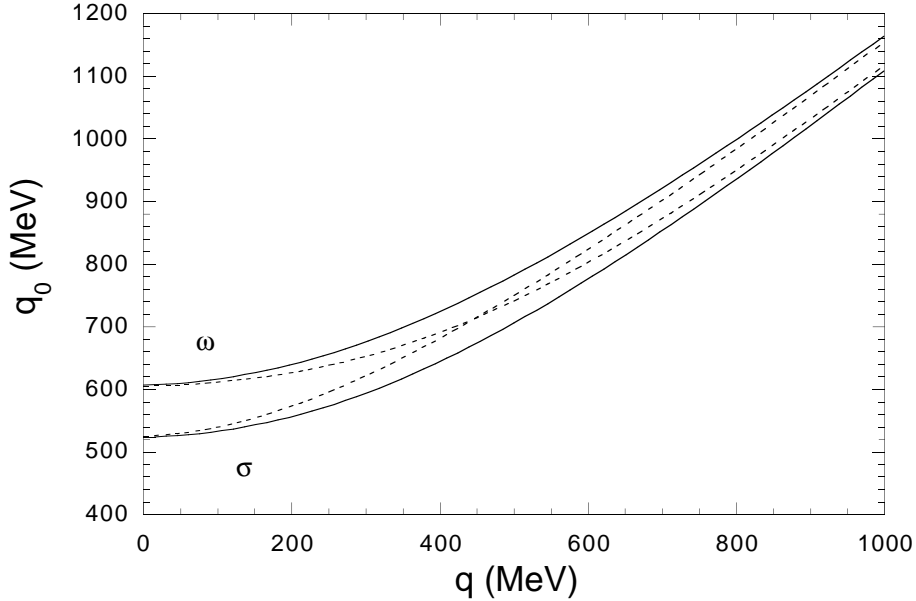,height=8cm}
\caption{Dispersion relation for the scalar and longitudinal modes at 
$\rho_B/\rho_0=2$.  The curves specified by $\omega$ are for the 
longitudinal component of the $\omega$ meson.  The dotted curves 
are the results without $\sigma$-$\omega$ mixing.  The solid 
curves include the effect of mixing.  (Note that the transverse mode is very 
close to the upper solid curve.)
 }
\label{f:disp}
\end{center}
\end{figure}

Now we are in a position to show our results.  First, 
in Fig.~\ref{f:disp}, we show the dispersion relation for the meson 
branches, which are calculated by searching for the zeros of the real part 
of the dielectric function given in Eq.(\ref{long}).  
As illustrated in the figure, 
if the $\sigma$-$\omega$ mixing is ignored, the longitudinal (L) 
mode and the $\sigma$ 
(S) mode cross each other at one point.  However, once the mixing 
is involved, the two modes {\em never\/} cross each other.  We can 
understand this phenomenon as a level-level repulsion due to the 
mixing, which is 
quite familiar in conventional nuclear physics, for example, in 
the Nilsson diagram (see Ref.\cite{greiner})).  
Adding the off-diagonal (or mixing) matrix element in the 
Hamiltonian leads to 
new eigenstates which never cross each other.  The new states 
approach the original crossing modes as the off-diagonal part becomes 
weaker.  This occurs at mid and high density (above $\sim \rho_0$) 
because its origin is 
the mixing effect, which vanishes at $\rho_B=0$.  

\begin{figure}[htb]
\begin{center}
\epsfig{file=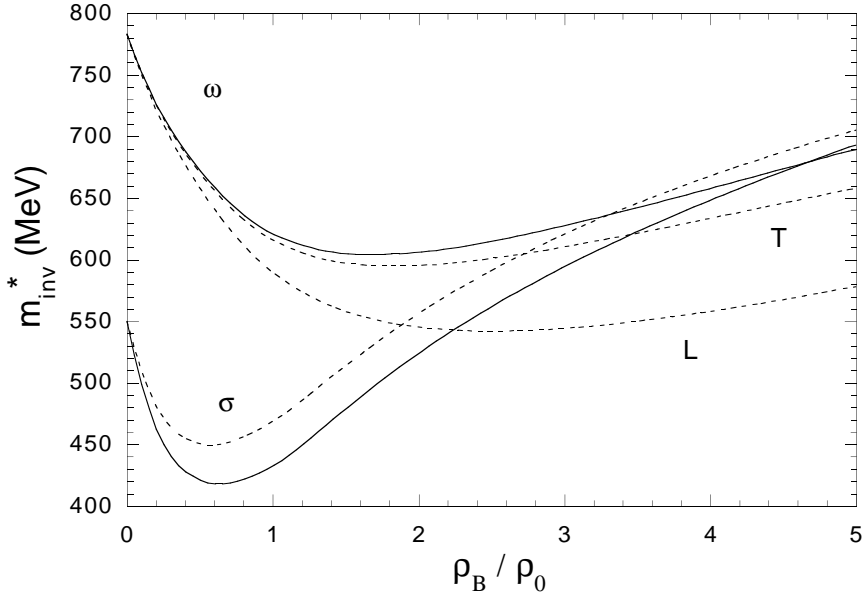,height=8cm}
\caption{``Invariant mass'' without $\sigma$-$\omega$ mixing.  
The solid curves 
are for three-momentum transfer $q$=1 MeV.  The L and T modes of the 
$\omega$ are almost degenerate.  The dotted curves are for $q$=500 
MeV, in which case the L and T modes are well separated.  
 }
\label{f:inv0}
\end{center}
\end{figure}
\begin{figure}[htb]
\begin{center}
\epsfig{file=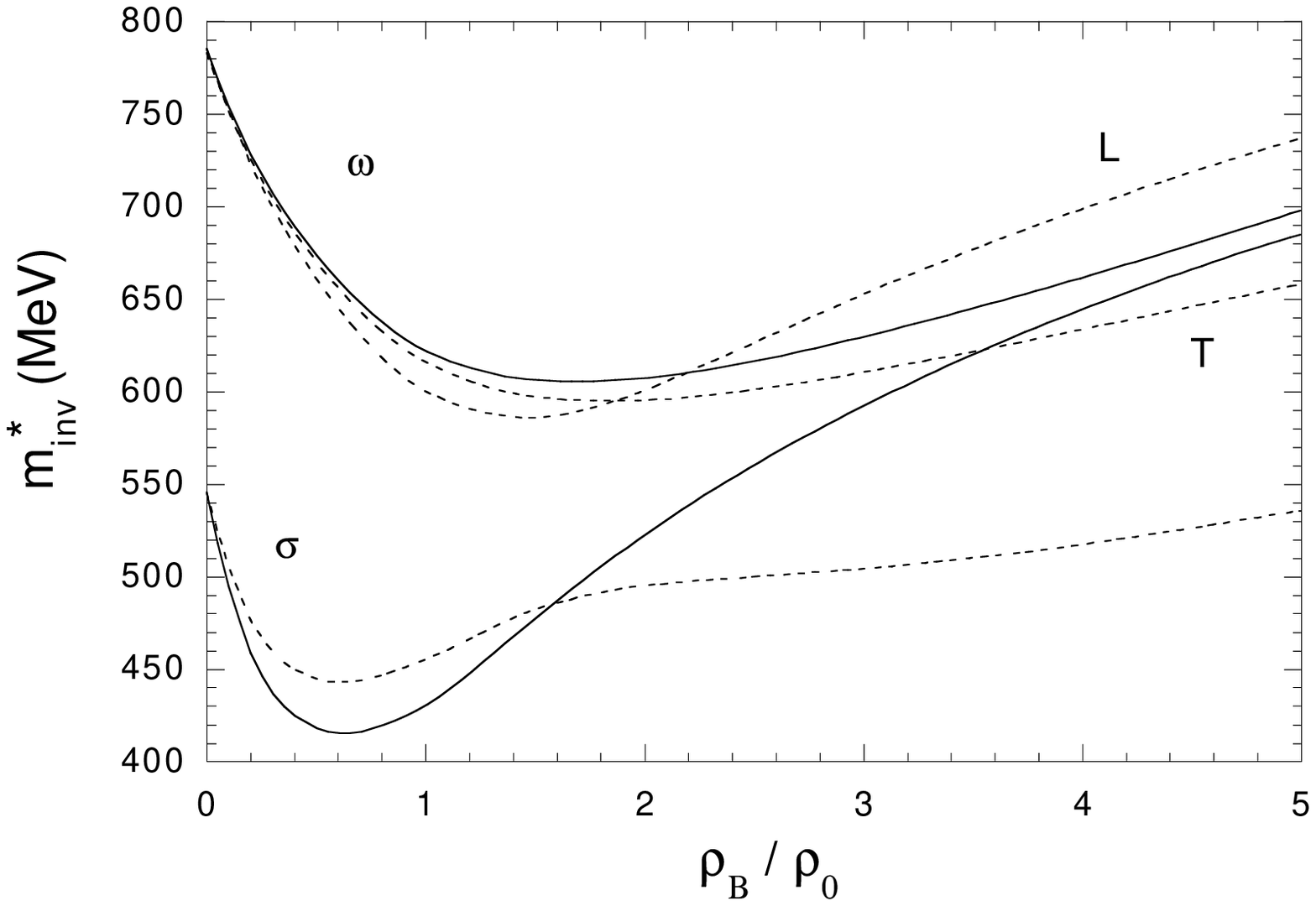,height=8cm}
\caption{``Invariant mass'' including $\sigma$-$\omega$ mixing.  
The solid curves
are for $q$=1 MeV.  The L and T modes of the
$\omega$ are again very close to each other.  The dotted curves are for 
$q$=500 MeV.  
 }
\label{f:invm}
\end{center}
\end{figure}
In Figs.~\ref{f:inv0} and \ref{f:invm}, we show the ``invariant mass'' 
($m_{inv}^* \equiv \sqrt{q_0^2 - {\vec q}^2}$, with $q_0$ chosen as in
Fig. 1 so that the real part of the dielectric function vanishes for
that value of $q$) as a function of density.
As shown in 
the figures, we can clearly see the role of mixing, which 
is quite important in determining the meson mass at mid and high 
density.  If the mixing is ignored, the masses of the L and S modes 
cross each other, like the dispersion relation shown in 
Fig.~\ref{f:disp}, and the L mass is below the other two at high 
density.  However, 
in the case where the mixing is included the L and S masses never cross 
each other.  At high density the L mass is pushed upwards, while the 
S mass is pulled downwards because of mixing (through the 
level-level repulsion).  From the figures we can see that the mixing 
effect becomes vital above $\sim \rho_0$.  

\begin{figure}[htb]
\begin{center}
\epsfig{file=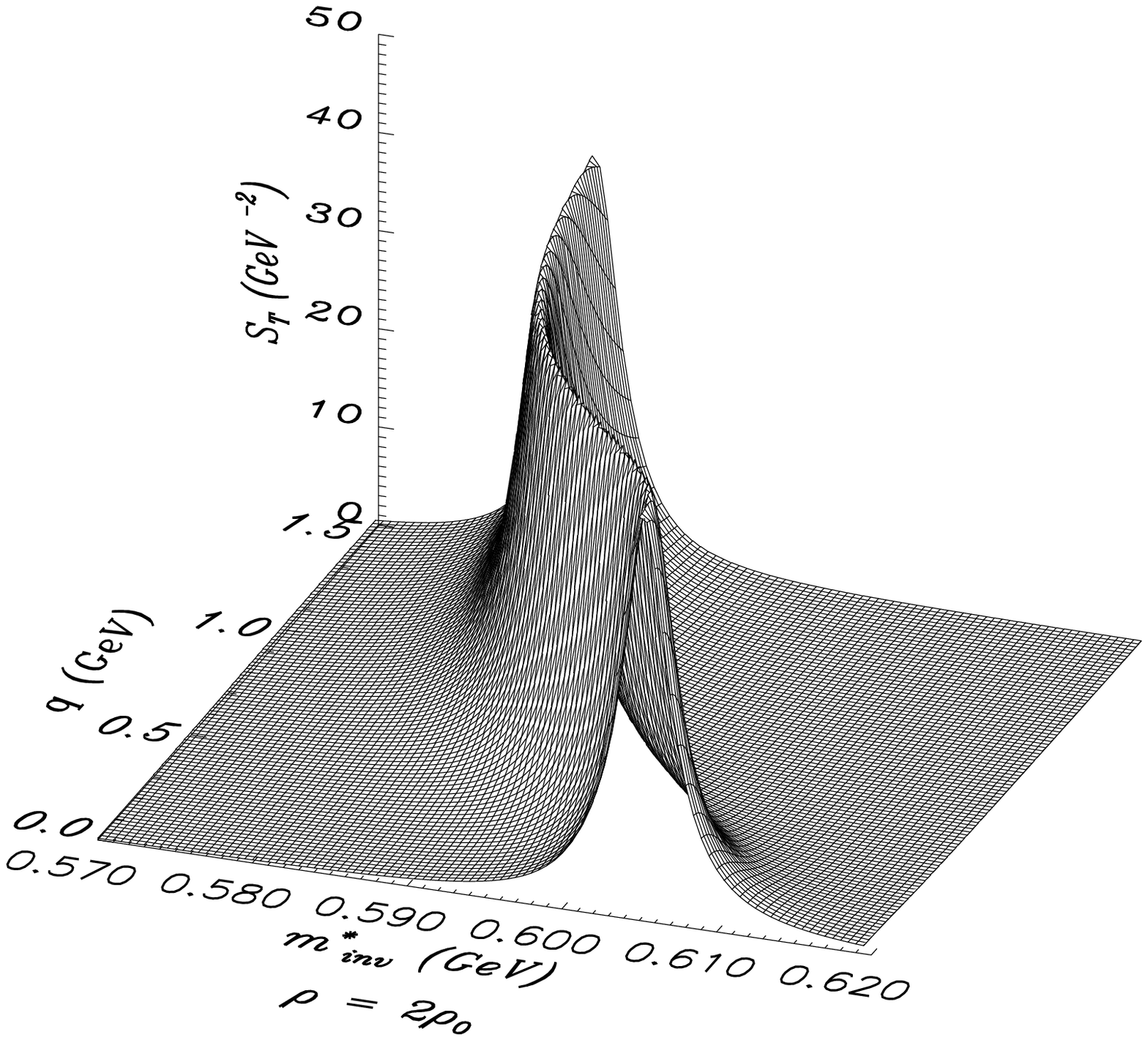,height=11cm}
\caption{Spectral function for the transverse $\omega$ at 
$\rho_B/\rho_0=2$. 
 }
\label{f:st}
\end{center}
\end{figure}
\begin{figure}[htb]
\begin{center}
\epsfig{file=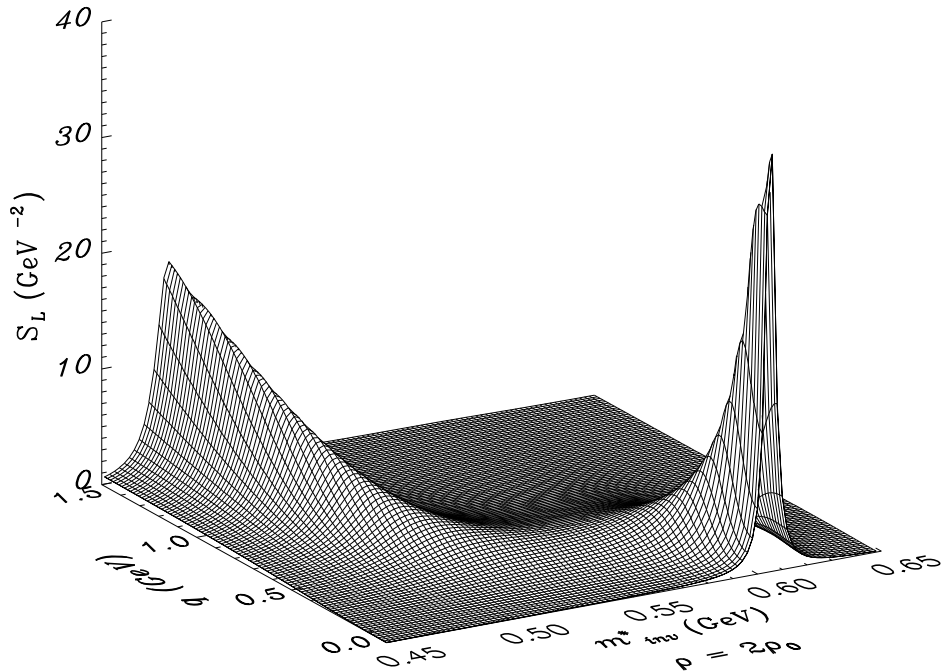,height=11cm}
\caption{Spectral function for the longitudinal $\omega$ at
$\rho_B/\rho_0=2$.
 }
\label{f:sl}
\end{center}
\end{figure}
\begin{figure}[hbt]
\begin{center}
\epsfig{file=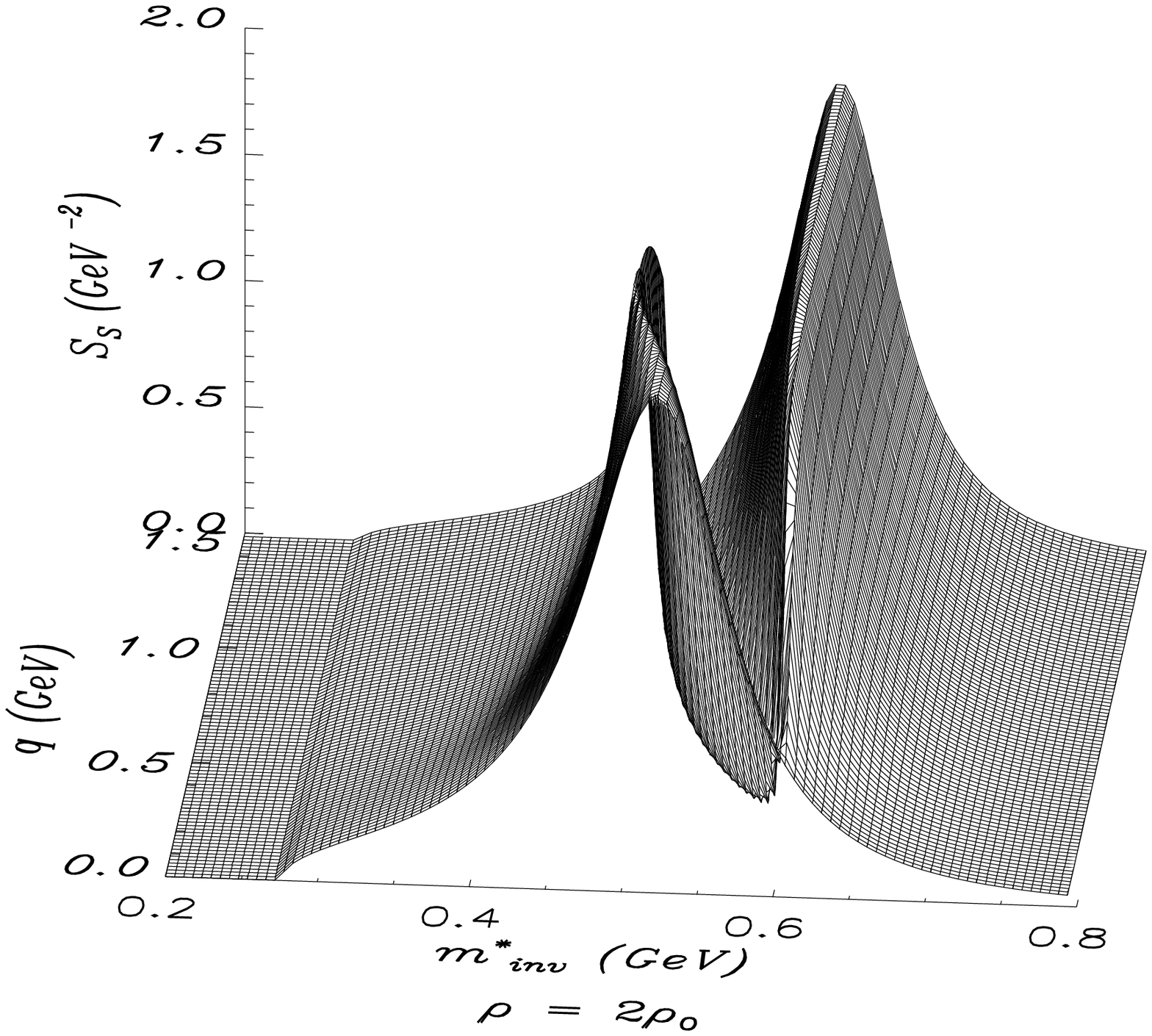,height=11cm}
\caption{Spectral function for the $\sigma$ at
$\rho_B/\rho_0=2$.
 }
\label{f:ss}
\end{center}
\end{figure}

Next we calculate the spectral functions of the T, L and S modes.  
It is very interesting to study these functions because in a thermal 
model the dilepton yields 
in heavy ion collisions would be proportional to the spectral 
functions. The spectral function is usually defined in terms of 
the imaginary part of the full propagator.  For the T mode 
it is:
\bge
S_T(m_{inv}^*,q,\rho_B) = - \frac{1}{\pi} {\Im}m\left[ 
\frac{d_0}{1 - d_0 \Pi_T} \right].   \label{spct} \\
\ene
While, in practice, it may be difficult or even impossible to 
separate the L and S modes in an experiment, it is of considerable
theoretical interest to study the following spectral functions:
\bg
S_L(m_{inv}^*,q,\rho_B) &=& - \frac{1}{\pi} {\Im}m\left[ 
\frac{d_0(1-\Delta_0\Pi_s)}{\epsilon_{SL}} \right], \label{spcl} \\
S_S(m_{inv}^*,q,\rho_B) &=& - \frac{1}{\pi} {\Im}m\left[
\frac{\Delta_0(1-d_0\Pi_L)}{\epsilon_{SL}} \right], \label{spcs} 
\en
corresponding to the complete, diagonal, longitudinal $\omega$ and
$\sigma$ propagators, respectively.

In Figs.~\ref{f:st}, \ref{f:sl} and \ref{f:ss}, we show the shape of 
the spectral function $S_i$ ($i$=T, L or S),  
as a function of the invariant mass and 
three-momentum transfer, at $\rho_B/\rho_0$=2.  For the T mode the shape is 
very simple, while for the L and S modes they are complicated 
because of the effect of mixing.  In particular, the S mode is quite 
remarkable: at $q$=0 MeV there is only one peak, while at fixed, 
finite $q$  there exist {\em two peaks\/} in the spectral function.  
Note that at $q$=0 MeV the mixed polarization insertion vanishes, 
so this two-peak structure is clearly associated with $\sigma$-$\omega$ 
mixing. 
As the density 
grows the two peaks becomes separated more widely.  In the L mode 
there is no such structure at fixed $q$.  However, the peak at 
low $q$ and that at high $q$ are separated very well and the values 
of their invariant masses are quite different.  As the density 
goes up this tendency becomes more clear.  

{}Finally, we should comment on the width of the $\sigma$ in free 
space.  In our calculation we have used $\Gamma_\sigma^0$=300 
MeV\cite{sigma} (see Eq.(\ref{swidth})), which is somewhat smaller 
than usual.  We have therefore also calculated the spectral 
functions using a width of 600 MeV but, as the results were not 
qualitatively different we do not report them here.  

In summary, we have studied the propagation of  $\sigma$ and 
$\omega$ mesons and the effect of $\sigma$-$\omega$ mixing 
in dense (symmetric) nuclear matter, within the Walecka model.  
We have illustrated that the effect of mixing is quite important in 
the propagation of the longitudinal $\omega$ and $\sigma$ mesons 
at a density above $\sim \rho_0$.  In the scalar (or $\sigma$) 
channel we have 
found a two-peak structure in the spectral function 
at finite three-momentum transfer, 
which could be measured in future experiments\cite{shim}.  
It would clearly be very 
interesting to compare these results with the medium modification of the  
meson propagation using the quark-meson coupling (QMC) model\cite{saito}, 
in which 
the effect of hadron internal structure is involved.  We will report 
such a study in the near future.

\vspace{1cm}
This work was supported by the Australian Research Council, and 
the Japan Society for the Promotion of Science.  
K.S. thanks T. Kunihiro, H. Shimizu and T. Hatsuda for fruitful 
comments and discussions.  
\clearpage
%
%

%

\begin{thebibliography}{99}
\bibitem{iv} The enhancement of the $\omega$ width due to the $\pi N$ 
channel in nuclear matter is expected to be rather small, for 
example, $\sim 30$ MeV at normal nuclear matter density 
(see Ref.\cite{fri}).
\end{thebibliography}
\end{document}